\documentclass{article}

\usepackage{microtype}
\usepackage{graphicx}
\usepackage{subfigure}
\usepackage{booktabs} % for professional tables
\usepackage{soul} % for \hl highlighting
\usepackage{amsfonts} % for \mathbb
\usepackage{amsmath}
\usepackage{algorithmic}
\usepackage{multirow}

\usepackage{hyperref}

\usepackage[accepted]{mlsys2026}

\makeatletter
\renewcommand{\Notice@String}{}
\makeatother

\mlsystitlerunning{VPP: Virtual Pipeline Parallelism for Efficient Chunked Prefill in Long-Context LLM Inference}

\begin{document}

\twocolumn[
\mlsystitle{VPP: Virtual Pipeline Parallelism for Efficient Chunked Prefill in Long-Context LLM Inference}

% unless you've provided the [accepted] option to the mlsys2026


\mlsyssetsymbol{equal}{*}
\mlsyssetsymbol{corr}{$\dagger$}

\begin{mlsysauthorlist}
\mlsysauthor{Yan Shi}{equal,hw,sjtu}
\mlsysauthor{Xiaochao Wang}{equal,hw}
\mlsysauthor{Jingchun Gao}{hw}
\mlsysauthor{Jintao Luo}{hw}
\mlsysauthor{Xinyi Zhou}{hw}
\mlsysauthor{Feng Liu}{hw}
\mlsysauthor{Kui Luo}{hw}
\mlsysauthor{Xushi Li}{corr,hw}
\mlsysauthor{Xinjie Guo}{hw}
\mlsysauthor{Liangjun Feng}{hw}
\end{mlsysauthorlist}

\mlsysaffiliation{hw}{Huawei Technologies Co., Ltd., Shanghai, China}
\mlsysaffiliation{sjtu}{Shanghai Jiao Tong University, Shanghai, China}

\mlsyscorrespondingauthor{Xushi Li}{lixushi@huawei.com}

\mlsyskeywords{Machine Learning, MLSys}

\vskip 0.3in

\begin{abstract}
    Chunked prefill pipeline parallelism (CPP) is a key technique for LLM inference. However, equal-size chunks exhibit imbalanced latency, as later chunks attend longer prefix KV caches and incur higher attention costs, leading to pipeline bubbles. Existing approaches mitigate this imbalance through dynamic chunk resizing (Dynamic CPP, DCPP), but our measurements show that this trades scheduling overhead for load balancing, which becomes unfavorable on long sequences. In this study, we propose Virtual Pipeline Parallelism (VPP), which keeps chunk sizes fixed and optimizes the pipeline layout through virtual stages. A V-shaped virtual-stage traversal overlaps each chunk's expensive middle stages with the lighter head and tail stages of its neighbors, while asynchronous communication and pipelined packing further reduce communication stalls and cross-request drain bubbles. We implement VPP in vLLM-Ascend and evaluate it on three MoE-based LLMs with sequences up to 1M tokens on 16 Ascend 910C NPUs. VPP improves throughput by up to 13.1\% over DCPP on long sequences and 6.7\% on mixed workloads, while preserving performance on short sequences. On a 512K-token DeepSeek-V3.1 prefill workload, VPP reduces the pipeline bubble ratio from 6.4\% to 0.1\%, achieving a 98.0\% reduction compared with DCPP.

\end{abstract}
]



\printAffiliationsAndNotice{*Equal contribution} 
% otherwise use the standard text.
% =========================================================
\section{Introduction}
\label{sec:Introduction}
% =========================================================

    % [Para 1 — LLM inference现状和变长序列Inference的挑战]
   The rapid development of large language model (LLM) applications \cite{openai2023gpt4, zhao2023survey}, particularly agent-based systems \cite{xi2023rise, wang2024survey, yao2023react} for multi-turn dialog \cite{ouyang2022instructgpt}, tool invocation \cite{schick2023toolformer}, knowledge retrieval \cite{lewis2020retrieval}, and complex reasoning \cite{wei2022chain}, has driven LLM inference workloads toward increasingly long and variable contexts \cite{xiang2026servegen, yuan2026agentic}. Serving sequences ranging from dozens to millions of tokens poses substantial challenges, including head-of-line (HOL) blocking, inefficient accelerator utilization, and growing memory pressure \cite{yu2022orca, kwon2023efficient, agrawal2023sarathi}.

    % [Para 2 — 把痛点收敛到 prefill，带出chunk prefill]
   These challenges are particularly pronounced during \emph{prefill}, which processes the input prompt in parallel and is generally compute-intensive. A long prefill can occupy accelerators for an extended period, delaying ongoing decodes and short requests queued behind it, increasing time-to-first-token (TTFT), and introducing inter-token latency (ITL) jitter. Chunked prefill \cite{agrawal2023sarathi} mitigates this problem by dividing a long prompt into smaller scheduling units that can be interleaved with decoding and other requests. Systems such as Sarathi \cite{agrawal2023sarathi} and Medha \cite{agrawal2024medha} exploit this finer granularity to improve utilization and alleviate HOL blocking.

   As model size and context length continue to grow, scheduling alone is insufficient, and multi-device parallelism becomes essential. Among existing parallel strategies, pipeline parallelism (PP) \cite{huang2019gpipe} is particularly attractive for large-scale deployment because it partitions the model by layers and communicates only intermediate activations between neighboring stages, requiring less frequent communication than tensor parallelism (TP) \cite{shoeybi2019megatron} and sequence parallelism (SP) \cite{korthikanti2023reducing}. Combining chunked prefill with PP leads to \emph{Chunked Prefill Pipeline Parallelism} (CPP), where prefill chunks become the fundamental work units flowing through pipeline stages. However, equal-sized chunking in CPP leads to progressively higher attention costs for later chunks due to their longer prefix KV caches, resulting in increasing chunk latency along the sequence, which in turn causes uneven pipeline progress and substantial bubbles. 
   
   Existing approaches mitigate this imbalance by dynamically adjusting chunk boundaries to equalize execution times \cite{li2021terapipe, sun2024seq1f1b, sglang_pp_blog, vllm_ascend_cpp_design}, a paradigm we refer to as \emph{Dynamic Chunked Prefill Pipeline Parallelism} (DCPP). However, dynamic resizing introduces additional system complexity, requiring accurate cost estimation, runtime calibration, and careful parameter tuning. More importantly, it fragments execution into finer-grained units. Our measurements show that, as sequence length grows, the resulting computation and communication overhead can eventually outweigh the benefit of bubble reduction.

   Our key insight is that the heterogeneous latency of fixed-size prefill chunks is not arbitrary. As the prefix context grows, successive chunks exhibit an approximately linear increase in execution latency. Instead of eliminating this heterogeneity through dynamic resizing, we exploit its predictable growth to construct a balanced pipeline schedule. Based on this insight, we propose \emph{Virtual Pipeline Parallelism} (VPP), which keeps chunk sizes fixed and reshapes the pipeline layout through virtual stages. A V-shaped stage traversal interleaves neighboring chunks such that the heavier middle stages of one chunk are paired with the lighter head and tail stages of adjacent chunks, absorbing the increasing chunk cost into parallel execution and substantially reducing imbalance-induced bubbles.

   To further reduce communication stalls and pipeline drain bubbles, VPP incorporates two complementary optimizations. First, asynchronous communication with cross-chunk stage reordering improves communication-computation overlap and reduces exposed communication latency. Second, pipelined packing fills otherwise idle drain periods with the leading chunks of the next request, reducing cross-request bubbles. Together, these mechanisms address computation imbalance, exposed communication, and cross-request idle time without dynamically resizing chunks.

    % [Para 9 — 评估目标与实验设置]
    We implement VPP in vLLM-Ascend\footnote{The implementation of VPP is open-sourced at \url{https://github.com/RookieCoder-Camera/vllm-ascend/tree/vpp-dev}.} and evaluate it on three representative MoE-based LLMs---Qwen3, DeepSeek-V3.1, and GLM-5.2---using 16 Huawei Ascend 910C NPUs. Our evaluation covers short (4K--16K), long (64K--1M), and mixed-length workloads, and includes detailed performance breakdowns, ablation studies, and chunk-size sensitivity analysis. In summary, the main contributions of our paper include:
    \begin{itemize}
        \item We quantify the limitations of Dynamic Chunked prefill Pipeline Parallelism (DCPP), showing that the execution fragmentation introduced by dynamic chunk resizing can outweigh its load-balancing benefit on long sequences.
        \item We propose and implement VPP on vLLM-Ascend. Instead of eliminating unequal chunk latency through dynamic resizing, VPP exploits its predictable growth to construct a balanced pipeline schedule. It keeps chunk sizes fixed and reshapes the pipeline layout through V-shaped stage traversal, asynchronous communication reordering, and pipelined packing.
        \item We evaluate VPP on three MoE models across short, long, and mixed workloads, demonstrating up to 13.1\% (long) and 6.7\% (mixed) throughput gains over DCPP while preserving short-sequence performance. On a 512K-token DeepSeek-V3.1 prefill workload, VPP reduces the pipeline bubble ratio from 6.4\% to 0.1\%, achieving a 98.0\% reduction compared with DCPP.
    \end{itemize}

% =========================================================
\section{Background}
\label{sec:Background}
% =========================================================

\subsection{Transformer Architecture}
\label{sec:Background.Transformer}
    
    LLMs are built upon the Transformer \cite{vaswani2017attention}, stacking multi-head self-attention and feed-forward networks (FFN) with layer norm and residual connections, typically in a decoder-only configuration \cite{radford2018improving, radford2019language, brown2020language}. The self-attention mechanism projects inputs into queries, keys, and values \(Q,K,V\in\mathbb{R}^{L\times d}\), and computes
    \begin{equation}
        \mathrm{Attn}(Q,K,V) = \mathrm{softmax}(QK^\top/\sqrt{d})V.    \end{equation}
    The resulting attention computation scales quadratically with sequence length, with a complexity of \(O(L^2d)\). In decoder-only Transformers, self-attention is causal, such that each token attends only to itself and preceding tokens.

    The FFN applies token-wise nonlinear transformations and typically accounts for a substantial fraction of the model parameters and computation. Many recent LLMs further adopt Mixture-of-Experts (MoE) architectures \cite{shazeer2017outrageouslylargeneuralnetworks, fedus2022switchtransformersscalingtrillion}, where the dense FFN is replaced by multiple expert networks and a router selectively activates a subset of experts for each token. This sparse activation increases model capacity without proportionally increasing computation, while introducing additional expert routing and communication.

\subsection{LLM Inference}
\label{sec:Background.LLMInference}

    Serving an LLM request consists of two phases with distinct computational characteristics. \emph{Prefill} processes all prompt tokens in parallel to generate the first output token and populate the KV cache \cite{shazeer2019fasttransformerdecodingwritehead}. Since many tokens are processed simultaneously, prefill is generally compute-bound. In contrast, \emph{Decode} generates tokens autoregressively, where each step processes a single token but accesses the accumulated KV cache, making it memory-bound.

    These phases are measured by different latency metrics. \emph{Time-to-first-token} (TTFT) captures the delay before the first token and is dominated by prefill execution, while \emph{inter-token latency} (ITL) measures the delay between generated tokens during decode. Due to the compute-intensive nature of prefill, long prompts can occupy accelerators for extended periods, increasing TTFT and interfering with concurrent requests.

\subsection{Parallelism Strategies}
\label{sec:Background.Parallelism}

    When a single GPU can no longer meet the throughput, latency, or memory demands of a model, computation must be distributed across multiple devices. \emph{Data parallelism} (DP) replicates the model and shards the input across GPUs. \emph{Tensor parallelism} (TP) \cite{shoeybi2019megatron} partitions weight matrices across GPUs along rows or columns, incurring collective communication at each layer. \emph{Context parallelism} (CP) \cite{yang2025contextparallelismscalablemilliontoken} partitions long input sequences and distributes KV caches across devices to reduce per-device memory footprint, at the cost of additional communication. 

    \emph{Pipeline parallelism} (PP) \cite{huang2019gpipe} instead partitions the model by layers. Each device hosts a consecutive group of layers, referred to as a pipeline stage, and intermediate activations are transferred between neighboring stages. Compared with TP and CP, PP communicates only stage boundary activations and therefore is attractive for large-scale long-context serving.

    However, PP suffers from pipeline bubbles, where stages become idle due to limited parallelism or uneven execution progress. These bubbles mainly originate from two sources. First, \emph{pipeline fill and drain}: stage dependencies prevent all stages from being simultaneously utilized during the warm-up and cool-down phases. Second, \emph{execution imbalance}: different chunks or micro-batches may incur different processing times, causing faster stages to wait for slower ones. While fill-and-drain bubbles are inherent to pipeline execution, imbalance-induced bubbles can be reduced through better scheduling.

\subsection{Chunked Prefill Pipeline Parallelism}
\label{sec:Background.CPP}

    \emph{Chunked prefill} \cite{agrawal2023sarathi} splits a long prompt into smaller chunks so that decoding and short requests can be interleaved between them, mitigating head-of-line blocking. When combined with pipeline parallelism, these prefill chunks become the work units that flow through the stages, which we refer to as \emph{Chunked prefill Pipeline Parallelism} (CPP).

    Although chunking improves scheduling flexibility, equal-size chunks introduce execution imbalance in long-context inference. Under causal attention, later chunks process longer prefix KV caches and therefore require more computation. Consequently, chunk latency increases with the chunk index, causing uneven stage utilization and pipeline bubbles.

    A natural approach is to dynamically adjust chunk sizes such that different chunks have similar execution times, which we refer to as \emph{Dynamic Chunked prefill Pipeline Parallelism} (DCPP). Instead of keeping a fixed token budget, DCPP estimates the execution cost of future chunks and modifies chunk boundaries accordingly.

% =========================================================
\section{Related Work}
\label{sec:RelatedWork}
% =========================================================

\subsection{Pipeline Parallelism Optimization}

    Pipeline parallelism has been extensively optimized through fine-grained execution and scheduling to reduce pipeline bubbles. GPipe \cite{huang2019gpipe} introduced micro-batching to improve pipeline utilization, while subsequent systems, including PipeDream \cite{narayanan2019pipedream}, Hanayo \cite{Liu_2023}, and Zero Bubble Pipeline Parallelism \cite{qi2024zero}, explored more flexible pipeline schedules, stage partitioning, and computation reordering to further reduce idle time during training.

    Pipeline load balancing has also been explored for LLM serving. gLLM \cite{guo2025gllmglobalbalancedpipeline} employs token throttling to balance computation across pipeline micro-batches based primarily on token counts,  while context-dependent attention costs are not explicitly modeled.

\subsection{Dynamic Chunk Scheduling}

    Dynamic sequence partitioning has also been explored to address pipeline imbalance. In training, TeraPipe \cite{li2021terapipe} formulates token-level sequence partitioning as a dynamic programming problem to search for non-uniform split points, while Seq1F1B \cite{sun2024seq1f1b} estimates chunk computation using theoretical FLOPs and adjusts chunk boundaries to balance execution time.
    
    More recently, dynamic chunk scheduling has been adopted in LLM serving systems. SGLang \cite{sglang_pp_blog} and vLLM-Ascend \cite{vllm_ascend_cpp_design} employ runtime profiling and online calibration to predict chunk latency and dynamically adjust chunk sizes. These runtime-aware approaches improve load balance under varying prefix lengths, but require online cost estimation and dynamic adaptation of chunk boundaries.

% =========================================================
\section{Motivation}
\label{sec:Motivation}
% =========================================================

    \begin{figure*}[t]
      \includegraphics[width=\linewidth]{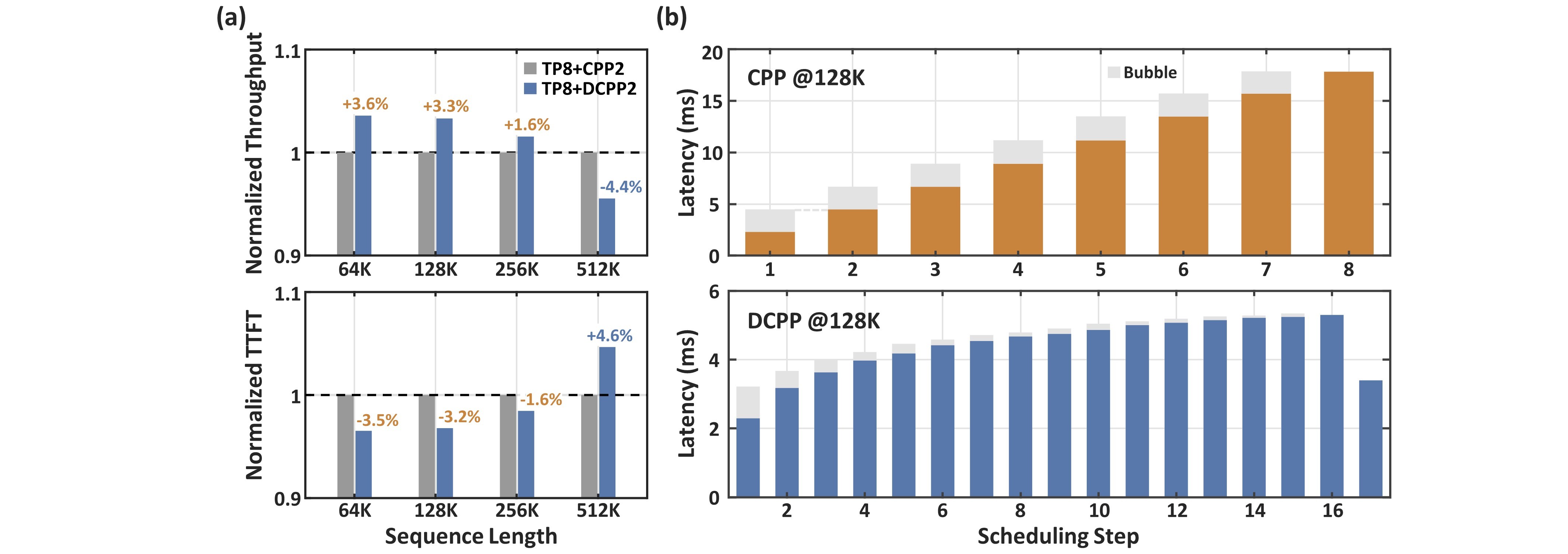}
      \caption{Performance of DCPP against CPP. (a) Normalized throughput and TTFT on long sequences, with TP8+CPP2 as the baseline. Labels above TP8+DCPP2 indicate its gain over the other, where orange labels denote gains, blue losses. (b) Per-chunk latency of both strategies on a 128K sequence with 32K chunk size; gray regions denote bubbles.}
      \label{fig:CPPandDCPPvsCPP}
    \end{figure*}

    \begin{table}[t]
    \centering
    \small
    \caption{Breakdown of DCPP's TTFT degradation over CPP on the 512K DeepSeek-V3.1 prefill workload. $\Delta = \text{DCPP} - \text{CPP}$; positive values denote degradation.}
    \label{tab:CPPandDCPPBreakdown}
    \vskip 0.1in
    \begin{tabular}{lrrrr}
        \toprule
         & \multicolumn{1}{c}{CPP (s)} & \multicolumn{1}{c}{DCPP (s)} & \multicolumn{1}{c}{$\Delta$ (s)} & \multicolumn{1}{c}{$\Delta$ / CPP} \\
        \midrule
        Computing          & 252.59 & 279.07 & +26.48 & +10.48\% \\
        Bubble & 38.64 & 20.47 & -18.17 & -47.03\% \\
        Exposed Comm. & 16.67 & 22.68 & +6.01 & +36.05\% \\
        \midrule
        \textbf{Total} & \textbf{307.90} & \textbf{322.22} & \textbf{+14.32} & \textbf{+4.65\%} \\
        \bottomrule
    \end{tabular}
    \end{table}

    We motivate our design by comparing DCPP against CPP on DeepSeek-V3.1 prefill across sequence lengths from 64K to 512K, using TP8+CPP2 and TP8+DCPP2. For each configuration, we sweep the chunk-size budget over \{8K, 16K, 24K, 32K\} and report the performance on best budget, averaged over 4 runs. Figure~\ref{fig:CPPandDCPPvsCPP}(a) reports the end-to-end performance. DCPP outperforms CPP on 64K--256K sequences, delivering consistent 1.6--3.6\% improvements in both throughput and TTFT. However, its advantage progressively diminishes with sequence length and reverses at 512K, resulting in a 4.4\% throughput loss and a 4.6\% TTFT increase.

    To understand this tradeoff, Figure~\ref{fig:CPPandDCPPvsCPP}(b) compares the per-chunk scheduling of two schemes on a 128K prompt with 32K chunk size; shaded gray regions indicate bubbles between consecutive chunks. CPP completes the prompt in 80.5\,s across 8 scheduling steps. Its per-chunk latency grows approximately linearly with the chunk index, as later chunks attend to increasingly larger KV caches, leading to substantial load imbalance. DCPP mitigates this imbalance by dynamically resizing chunks to equalize their execution times, substantially reducing pipeline bubbles. However, imperfect runtime estimation prevents complete balancing. DCPP completes the same prompt in 74.9\,s, 7.0\% faster than CPP, but requires 17 scheduling steps---more than twice CPP's 8---resulting in finer-grained execution and reduced operator efficiency.

    This fragmentation overhead increasingly offsets DCPP's load-balancing benefit as sequence length grows. At 512K, where the optimal chunk-size budget is 24K, profiler traces show that CPP requires 22 invocations, whereas DCPP requires 97, or $4.4\times$ as many. Table~\ref{tab:CPPandDCPPBreakdown} decomposes the resulting TTFT. DCPP's load-balancing mechanism remains effective: it reduces bubble time by 47.0\%, saving 18.17\,s and halving the bubble ratio from 12.5\% to 6.4\%. However, these savings are outweighed by the cost of finer-grained execution. Computation time increases by 10.5\% (+26.48\,s) as smaller chunks reduce operator efficiency, while the increased number of chunk boundaries raises exposed communication by 36.1\% (+6.01\,s). Despite dynamic balancing, 20.47\,s of bubble remains, and DCPP ultimately incurs a 14.32\,s (4.6\%) TTFT regression over CPP.

    These observations yield a critical insight: \textit{DCPP trades scheduling overhead for load-balancing gains.} As sequence length grows, however, DCPP requires progressively more invocations, and the resulting fragmentation overhead eventually outweighs the benefits of bubble reduction. Moreover, imperfect runtime estimation leaves residual bubbles even after dynamic balancing. These limitations motivate our design of VPP, which keeps chunk sizes fixed and instead reshapes the pipeline layout to accommodate unequal chunk latencies.

% =========================================================
\section{VPP: Design and Implementation}
% =========================================================

    \begin{figure*}[t]
      \includegraphics[width=\linewidth]{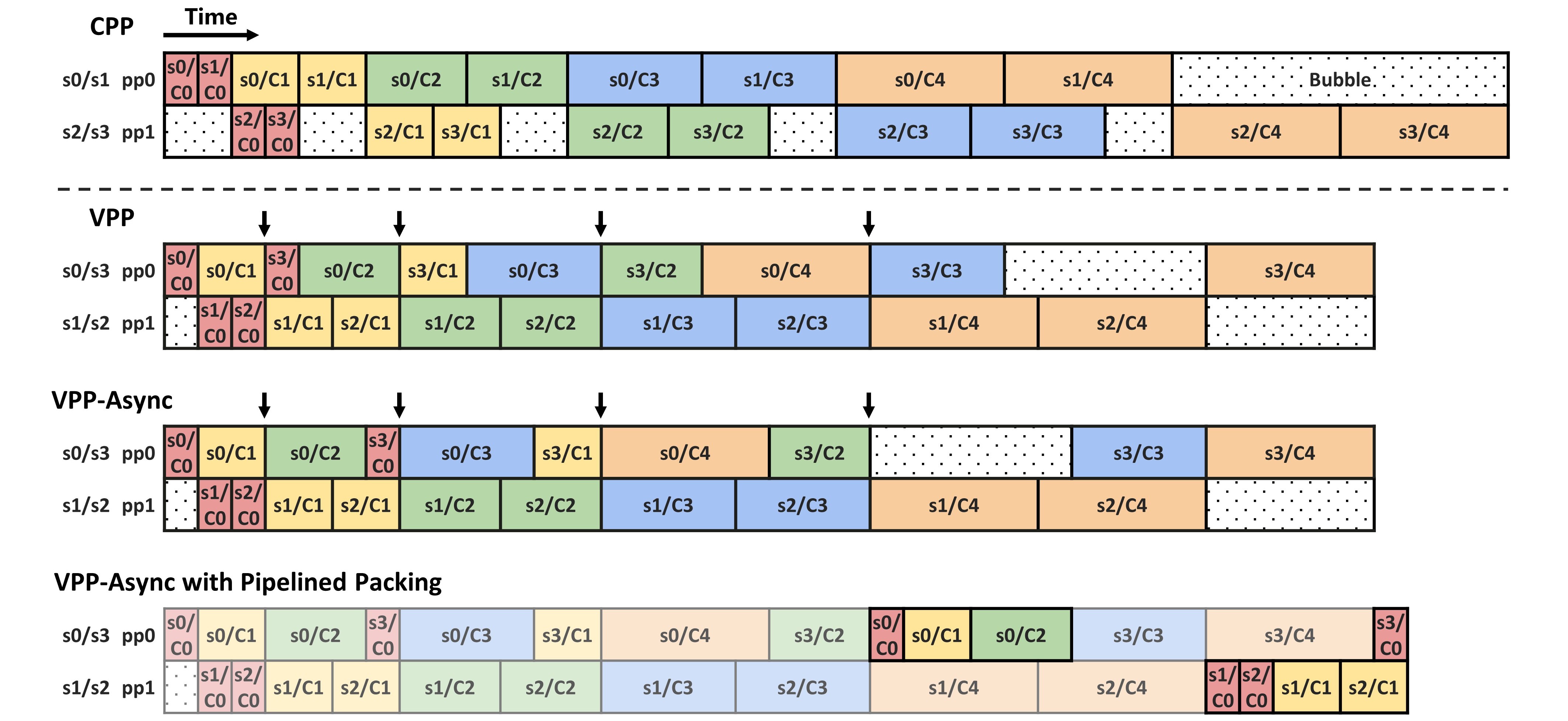}
      \caption{Comparison of scheduling timelines across CPP and the three VPP-based schemes. CPP exhibits linearly growing chunk latencies and pipeline bubbles, while VPP reduces compute idle via V-shaped traversal, VPP-Async hides communication stalls via reordering, and VPP-Async with Pipelined Packing further compresses drain bubbles by packing leading chunks of the next request into the tail window.}
      \label{fig:VPPFramework}
    \end{figure*}

    The preceding analysis shows that, while dynamic resizing mitigates CPP's load imbalance, its scheduling overhead grows substantially on long sequences. We argue that this imbalance is fundamentally a \emph{pipeline layout} problem rather than a chunk-size problem. Under causal attention, per-chunk latency grows approximately linearly with the chunk index (Figure~\ref{fig:CPPandDCPPvsCPP}(b)). In our EP-enabled configuration, the MoE-related fixed overhead is relatively small, making the prefix-dependent attention cost the dominant source of this latency growth. 
    
    This regularity exposes an opportunity for a fixed pipeline layout. Rather than adapting chunk sizes to fit the pipeline, VPP keeps chunk sizes fixed and partitions the model into multiple virtual stages, which are mapped onto physical pipeline ranks in a V-shaped fold-back layout to accommodate unequal chunk latencies. Figure~\ref{fig:VPPFramework} illustrates three progressively optimized schemes: VPP, VPP-Async, and VPP-Async with Pipelined Packing. They reduce idle time through V-shaped stage traversal, asynchronous communication, and cross-request packing, respectively.

\subsection{VPP: V-shaped pipeline scheduling}
\label{sec:VPP.VPP}
    
    We consider a PP configuration with two physical ranks, $pp_0$ and $pp_1$. The model is partitioned into four virtual stages $s_0, s_1, s_2, s_3$, with boundary stages placed on $pp_0$ ($s_0, s_3$) and middle stages on $pp_1$ ($s_1, s_2$). A long prompt is chunked into $N$ parts, denoted as $C_0, C_1, \dots, C_{N-1}$, with $N=5$ in the example shown in Figure~\ref{fig:VPPFramework}.
    
    This fold-back placement allows each chunk to traverse both ranks and return, rather than flowing unidirectionally as in conventional PP. Each chunk $C_k$ therefore follows a \emph{V-shaped} route:
    \[
      s_0\,(pp_0)\ \rightarrow\ s_1\,(pp_1)\ \rightarrow\ s_2\,(pp_1)\ \rightarrow\ s_3\,(pp_0).
    \]
    Specifically, $pp_0$ starts the chunk at $s_0$ and sends its activation to $pp_1$, which executes the two middle stages, $s_1$ and $s_2$, before sending the activation back to $pp_0$ for the final stage $s_3$. This forward-and-back traversal forms the characteristic ``V'' shape.

    The key to reducing pipeline bubbles is to exploit the near-linear growth of fixed-size chunk latency observed in Figure~\ref{fig:CPPandDCPPvsCPP}(b). Based on this empirical regularity, we approximate the per-stage latency of $C_k$ as
    \begin{equation}
        \tau_k = (k+1)t,
    \end{equation}
    where $t$ denotes the per-stage latency of the first chunk. While $pp_1$ executes the two middle stages of $C_k$, its execution time is approximately $2\tau_k = 2(k+1)t$. During the same interval, $pp_0$ can execute two ready operations: the \emph{exit} of the previous chunk, $s_3/C_{k-1}$, with latency $\tau_{k-1} = kt$, and the \emph{entry} of the next chunk, $s_0/C_{k+1}$, with latency $\tau_{k+1} = (k+2)t$. Their combined latency is
    \begin{equation}
        \tau_{k-1}+\tau_{k+1} = kt+(k+2)t =2(k+1)t = 2\tau_k.
    \end{equation}
    Thus, the work scheduled on $pp_0$ closely matches the middle-stage execution of $C_k$ on $pp_1$, allowing useful computation from neighboring chunks to fill the idle window caused by unequal chunk latencies. The leading chunk $C_0$ is the boundary case: while $pp_1$ executes its two middle stages with latency $2\tau_0=2t$, $pp_0$ executes $s_0/C_1$ with latency $\tau_1 = 2t$, providing the same approximate match.
    
    This balance relies on the linear-$\tau_k$ regime, which holds when attention dominates and MoE is expert-parallel. If attention becomes less dominant or EP is disabled, the equality may no longer hold, causing steady-state bubbles to reappear. However, two inefficiencies remain: the bidirectional cross-rank handoffs at the arrows in Figure~\ref{fig:VPPFramework} are synchronous and blocking, introducing communication windows that cannot be overlapped; and the final chunks leave an unavoidable drain bubble at the tail. The next two schemes address these inefficiencies, respectively.

\subsection{VPP-Async: communication-computation overlap}
\label{sec:VPP.VPP-Async}

    In Figure~\ref{fig:VPPFramework}, each arrow in VPP diagram denotes a synchronous PP send/recv: the receiver blocks until the activation arrives, placing communication latency directly on the critical path. At the handoff between $C_1$ and $C_2$, for instance, $pp_0$ sends $s_0/C_2$ forward to $pp_1$ while $pp_1$ sends $s_2/C_1$ back to $pp_0$. Both ranks must wait for the corresponding transfer to complete, resulting in a bidirectional communication stall.
    
    This stall arises from operation ordering rather than an inherent data dependency. Once the required prefix KV cache is available, $pp_0$ can safely reorder its local stage execution to avoid simultaneous bidirectional transfers. Concretely, we swap the tail stage of $C_{k-1}$ ($s_3/C_{k-1}$) and the head stage of $C_{k+1}$ ($s_0/C_{k+1}$). After reordering, each transfer is issued while the peer rank is performing useful computation, allowing communication to overlap with computation. VPP-Async thereby removes most exposed communication stalls without changing the V-shaped layout.

\subsection{VPP-Async with Pipelined Packing: cross-request bubble compression}
\label{sec:VPP.VPP-AsynPacking}

    VPP-Async hides synchronous communication stalls within a single request through communication-computation overlap. However, the tail of a long request inevitably introduces a drain bubble: as the final chunks $C_{N-2}$ and $C_{N-1}$ complete their V-traversals, one rank runs out of useful work and remains idle until the request finishes. This tail inefficiency is inherent to pipelined execution when only a single request is available.

    In practical serving systems, requests are continuously queued and scheduled. The drain window of a request $R$ can therefore be reused to process the initial chunks of a subsequent request $R'$. We exploit this opportunity by packing the leading chunks of $R'$ into the tail bubble of $R$, whose duration is approximately $(N+1)t$ in our latency model. Since the early chunks of $R'$ incur the lowest latency ($\tau \approx t$), they naturally fit into the shrinking tail windows, as illustrated in Figure~\ref{fig:VPPFramework}. Across a stream of requests, this cross-request packing overlaps the drain of one request with the pipeline fill of the next. Figure~\ref{fig:VPPFramework} shows the fully packed case; when the tail window is insufficient, packing still reduces the effective drain bubble.

\subsection{Implementation}
\label{sec:VPP.Implementation}

    We implement VPP on top of vLLM-Ascend using PyTorch and HCCL. Control-plane components, including the virtual-stage scheduler, fold-back layer assignment, and per-batch pipeline state machine, are implemented in Python, while optimized attention and MoE kernels are reused without modification.
    
    VPP adopts a V-shaped fold-back topology: the first half of virtual stages traverse forward across ranks and the second half traverse backward. Fold points, where two consecutive virtual stages are mapped to the same physical device, require no inter-device communication. The implementation also supports uneven layer partitioning and user-specified layer ranges for memory and compute balancing.

    We extend vLLM V1's engine with a dual-queue scheduler to support fold-back execution. Batches waiting for their initial virtual-stage traversal are maintained in the first queue. After reaching the fold point, batches migrate to the continuation queue for final-stage execution and sampling. The scheduler maintains one additional in-flight batch and uses asynchronous task handles to avoid blocking the scheduling loop. Two alternating HCCL communication groups exchange activations through point-to-point send/recv, avoiding deadlocks during bidirectional transfers. VPP is enabled by a single flag; when disabled, execution reverts to upstream vLLM.

% =========================================================
\section{Evaluation}
\label{sec:Evaluation}
% =========================================================
    \begin{figure*}[t]
        \includegraphics[width=\linewidth]{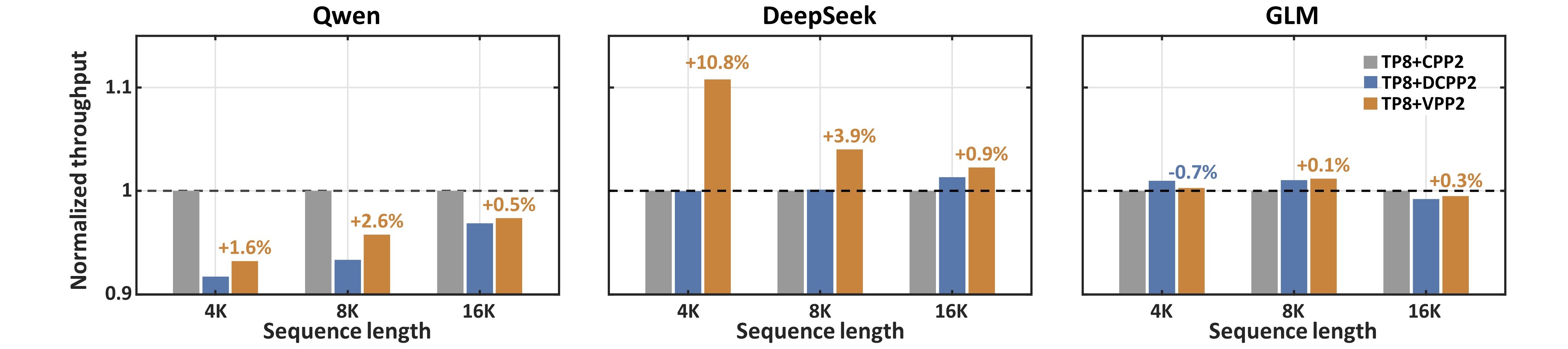}
        \caption{Normalized throughput on short sequences for all three models, with TP8+CPP2 as the baseline. Labels above TP8+VPP2 bars indicate its gain over TP8+DCPP2, where orange labels denote gains, blue losses.}
        \label{fig:E2EPerfShortSeq}
    \end{figure*} 

    In this section, we evaluate VPP against alternative pipeline parallelism strategies across diverse models, sequence lengths, and workload mixes. We focus on three questions:
    
    % TODO: including with prefix caching
    \begin{itemize}
    \item \textbf{Overall effectiveness:} How much performance improvement does VPP provide over existing pipeline parallelism schemes across diverse workloads?
    \item \textbf{Mechanism analysis:} Which sources of overhead does VPP eliminate or reduce according to profiler-based breakdowns?
    \item \textbf{Design sensitivity:} How do individual VPP components and chunk-size choices affect end-to-end performance?
    \end{itemize}
    
    Section~\ref{sec:Evaluation.Setup} describes our experimental setup. The subsequent three sections investigate each question in turn.

\subsection{Experimental Setup}
\label{sec:Evaluation.Setup}
    \textbf{Testbed.} 
    We conduct all experiments on a Huawei Atlas 900 A3 SuperPoD compute node equipped with 16 Ascend 910C NPUs (64\,GB HBM each), interconnected via HCCS. All collective communication uses HCCL. We use PyTorch v2.10.0 with torch\_npu, CANN v8.3, and vLLM v0.23.0 with vLLM-Ascend v0.23.0 as the inference framework; our VPP scheduler is implemented on top of vLLM-Ascend's pipeline parallelism infrastructure.
    
    \textbf{Models.} 
    We evaluate three representative MoE-based LLMs spanning different scales and architectural variants: \emph{Qwen3-Coder-30B-A3B-Instruct} (30.5B total, 3.3B active, 128 experts, GQA attention), \emph{DeepSeek-V3.1-Terminus} (671B total, 37B active, 256 experts, MLA attention), and \emph{GLM-5.2} (744B total, 40B active, 256 experts, MLA + DSA sparse attention). In the remainder of this paper, we use the abbreviations \emph{Qwen}, \emph{DeepSeek}, and \emph{GLM} to denote the above three models, respectively.
    
    \textbf{Parallelism configurations.} 
    We compare three strategies on 16 NPUs, all using 8-way tensor parallelism with two pipeline ranks: \emph{TP8+CPP2}, chunked prefill pipeline parallelism; \emph{TP8+DCPP2}, CPP with dynamic chunk sizing, configured with vLLM-Ascend's recommended settings (minimum chunk size: 4096 tokens, smoothing factor: 1.0); and \emph{TP8+VPP2}, our proposed scheme.
    
    \textbf{Workloads and metrics.} 
    We drive the serving system with AISBench under three workload suites: (1)~\emph{short sequence}, 100 concurrent requests with 4K, 8K, and 16K input tokens; (2)~\emph{long sequence}, one request at a time with 64K to 1M input tokens; and (3)~\emph{mixed-length sequence}, 500 requests synthesized from GSM8K with lengths spanning 7 to 122{,}710 tokens (mean $\approx$ 22.5K, median $\approx$ 10K, p90 $\approx$ 64K), served at concurrency~16. All requests generate one output token to isolate prefill performance. We report throughput (token/s) and TTFT (s). 
    
    \textbf{Methodology.} 
    Since chunked-prefill performance depends on the chunk size, we perform a sweep over $\{8K, 16K, 24K, 32K\}$ for all strategies and report their best-performing settings. Each (strategy, workload, budget) combination is repeated four times, and we report the mean throughput and TTFT. For detailed analysis, we collect PyTorch profiler traces and request-level scheduling timelines to attribute performance differences to computation, bubble, and exposed communication.

\subsection{End-to-End Performance}
\label{sec:Evaluation.End2End}

    \begin{figure*}[t]
      \includegraphics[width=\linewidth]{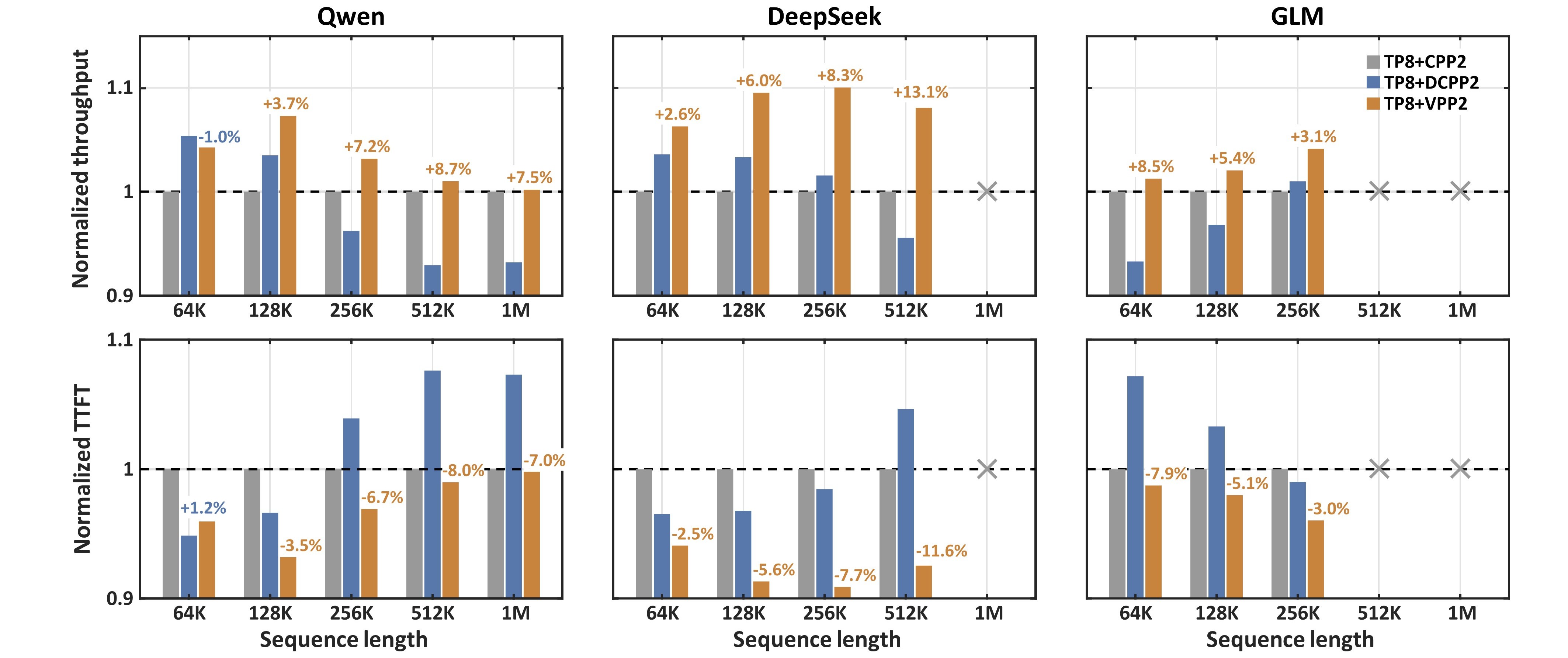}
      \caption{Normalized throughput and TTFT on long sequences for all three models, with TP8+CPP2 as the baseline. Labels above TP8+VPP2 bars indicate its gain over TP8+DCPP2, where orange labels denote gains, blue losses.}
      \label{fig:E2EPerfLongSeq}
    \end{figure*}

    \begin{figure}[t]
      \includegraphics[width=\linewidth]{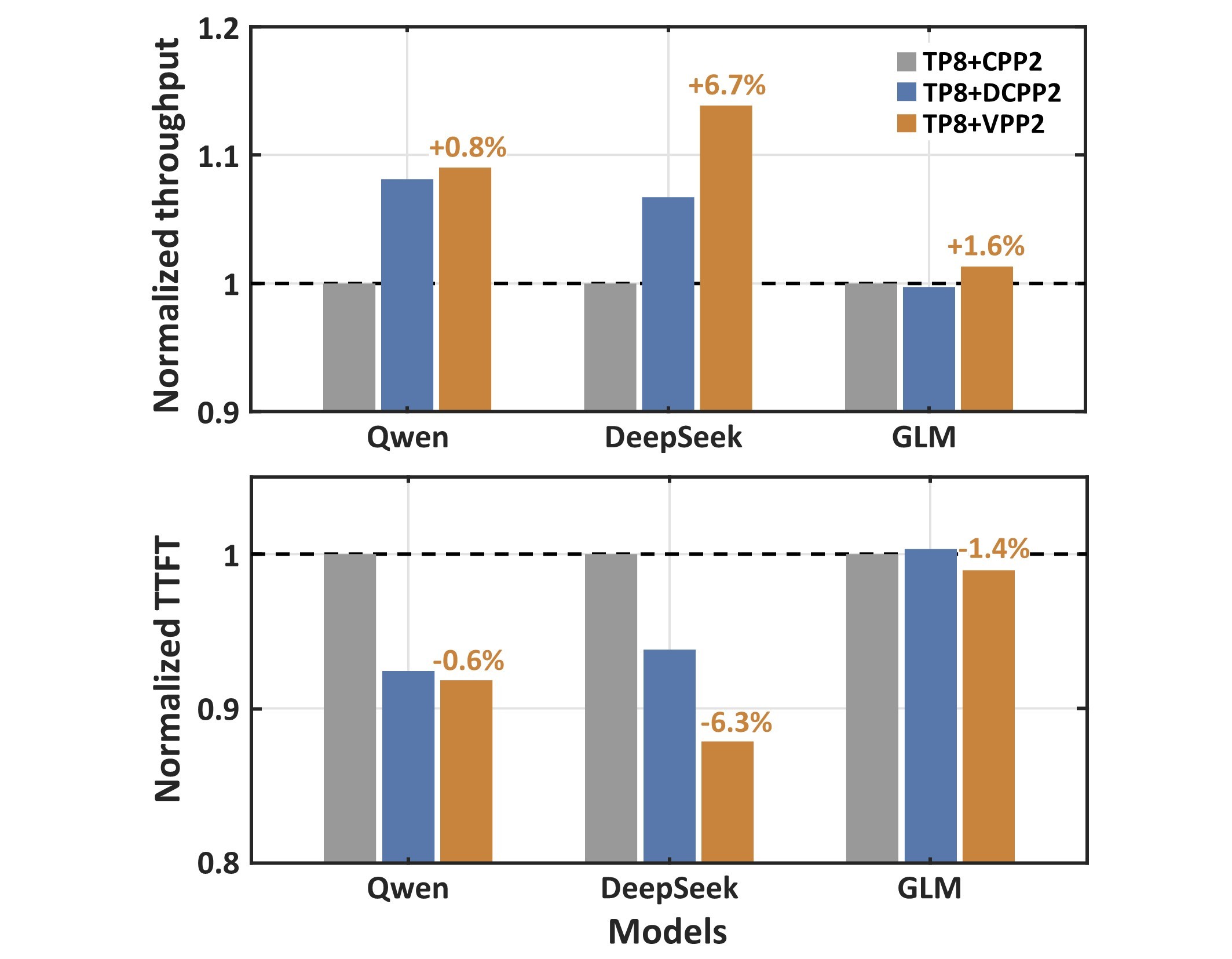}
      \caption{Normalized throughput and TTFT on the mixed-length workload for all three models, with TP8+CPP2 as the baseline. Labels above TP8+VPP2 bars indicate its gain over TP8+DCPP2, where orange labels denote gains, blue losses.}
      \label{fig:E2EPerfMixedSeq}
    \end{figure}

    We evaluate VPP under three representative serving scenarios: short sequences where pipeline bubbles are limited, long sequences where prefill computation dominates, and mixed-length workloads that resemble practical serving environments. We report each strategy's result at its optimal chunk size, normalized to TP8+CPP2 (1.0). The results below correspond to VPP-Async with pipelined packing, which is the full implementation of our proposed scheme. 

    \textbf{Short sequences.}
    Figure~\ref{fig:E2EPerfShortSeq} reports normalized throughput results on short sequences. We omit TTFT for short sequences because high concurrency causes queueing delay to dominate the measured latency, making it less representative of pipeline execution efficiency. VPP shows almost no regression against DCPP across all models: the two perform similarly on Qwen and GLM, while VPP achieves its largest advantage on DeepSeek, reaching 10.8\% at 4K chunk size. Compared with CPP, VPP improves performance on DeepSeek, achieves comparable results on GLM, and degrades by up to 6.8\% on Qwen. The degradation on Qwen is attributed to its smaller active parameter count (3.3B vs. 37B/40B), which makes scheduling overhead less amortizable at short sequence lengths. Overall, VPP preserves performance on short sequences, although the benefit of bubble elimination is limited when scheduling overhead dominates.

    \textbf{Long sequences.}
    Figure~\ref{fig:E2EPerfLongSeq} reports normalized throughput and TTFT on long sequences. VPP outperforms CPP and DCPP in nearly all configurations. Compared with CPP, it improves throughput by up to 7.3\% (Qwen), 10.0\% (DeepSeek), and 4.1\% (GLM), with comparable TTFT reductions. Compared with DCPP, VPP's advantage increases with sequence length on Qwen and DeepSeek, reaching 3.7\%--8.7\% on Qwen (128K--512K) and 2.6\%--13.1\% on DeepSeek (64K--512K). In contrast, on GLM, the gain peaks at 64K (8.5\%) and narrows at longer lengths. This difference correlates with the attention characteristics of each model (GQA/MLA vs. DSA), which we revisit in Section~\ref{sec:Evaluation.Operator}. These results confirm that VPP's interleaved scheduling is most effective on long sequences, where compute-intensive prefill chunks provide ample opportunity to hide pipeline bubbles and communication stalls.

    \textbf{Mixed sequences.}
    Figure~\ref{fig:E2EPerfMixedSeq} reports normalized throughput and TTFT on mixed-length workload. VPP improves upon both CPP and DCPP across all three models. On Qwen, VPP achieves throughput gains of 9.0\% and 0.8\% over CPP and DCPP, respectively, with similar TTFT reduction trends. On DeepSeek, VPP delivers the largest improvement, increasing throughput by 13.8\% over CPP and 6.7\% over DCPP. On GLM, VPP achieves more modest gains of 1.3\%--1.6\% over the other two strategies. These results demonstrate that VPP can effectively exploit fragmented idle slots in heterogeneous serving workloads through virtual-stage scheduling, while preserving the benefits of chunked prefill across different model architectures.

\subsection{Source of Gains: Performance breakdown}
\label{sec:Evaluation.Operator}
    
    To understand the sources of VPP's TTFT improvement, we profile the DeepSeek-V3.1 512K workload and decompose the performance difference into three dimensions: computing, pipeline bubbles, and exposed communication. Table~\ref{tab:VPPBreakdown} summarizes the contribution of each component. Overall, VPP reduces computation time by 10.1\% and non-computing overhead (bubbles and exposed communication) by 28.1\% compared with DCPP, resulting in a 12.53\% (40.36\,s) improvement in TTFT.
    
    % VPP breakdown
    \begin{table}[t]
    \centering
    \small
    \caption{Breakdown of VPP's TTFT improvement over DCPP on the 512K DeepSeek-V3.1 prefill workload. $\Delta = \text{VPP} - \text{DCPP}$; negative values denote improvement. Panel (a) decomposes the end-to-end latency; (b) and (c) detail the two largest contributors.}
    \label{tab:VPPBreakdown}
    \vskip 0.1in
    \begin{tabular}{lrrrr}
        \toprule
         & \multicolumn{1}{c}{DCPP (s)} & \multicolumn{1}{c}{VPP (s)} & \multicolumn{1}{c}{$\Delta$ (s)} & \multicolumn{1}{c}{$\Delta$/DCPP} \\
        \midrule
        \multicolumn{5}{l}{\textit{(a) End-to-end decomposition}} \\
        ~~Computation               & 279.07 & 250.84 & -28.23 & -10.11\% \\
        ~~Bubble         & 20.47  & 0.39   & -20.07 & -98.04\% \\
        ~~Exposed Comm.    & 22.68  & 30.63  & +7.95  & +35.05\% \\
        \midrule
        ~~\textbf{Net (profiler)} &        &        & \textbf{-40.36} & \textbf{-12.53\%} \\
        \midrule
        \multicolumn{5}{l}{\textit{(b) Computation detail}} \\
        ~~Attention$^{\star}$ & 228.13 & 233.08 & +4.96  & +2.17\% \\
        ~~Slice                    & 22.90  & 5.18   & -17.72 & -77.38\% \\
        ~~MatMulV3                 & 13.06  & 2.54   & -10.52 & -80.55\% \\
        ~~Other kernels  & 14.98  & 10.04  & -4.95  & -33.04\% \\
        \midrule
        \multicolumn{5}{l}{\textit{(c) Bubble detail}} \\
        ~~Free                     & 18.61  & 0.39   & -18.22 & -97.90\% \\
        ~~Launch stall             & 1.85   & 0.00036 & -1.85  & -99.98\% \\
        \bottomrule
    \end{tabular}
    \vskip 0.05in
    \parbox{\linewidth}{\footnotesize
    $^{\star}$FusedInferAttentionScore on Ascend NPU.
    }
    \end{table}

    \textbf{Computation.}
    Table~\ref{tab:VPPBreakdown}(b) compares execution granularity between VPP and DCPP. DCPP executes 4.8× more kernels and 4.7× more host launches than VPP due to its dynamically resized and fragmented execution schedule. The resulting operator overhead is largely avoided by VPP: Slice and MatMulV3 together account for 28.24\,s reduction, contributing nearly all of the computational savings. Interestingly, VPP's Attention kernel time is 4.96\,s longer than DCPP's, confirming that VPP's gain does not come from faster attention computation, but from reducing execution fragmentation around attention. This is because attention cost is dominated by the accumulated KV context: although VPP issues fewer attention invocations, each invocation processes a larger prefix context, limiting the reduction in total attention time.

    \textbf{Bubble.}
    Table~\ref{tab:VPPBreakdown}(c) breaks down the bubble time on the critical rank. In DCPP, the critical path lies on $pp_1$, which waits 1.85 s for the first chunk from $pp_0$. In VPP, the critical path shifts to $pp_0$, which starts immediately upon request arrival, reducing launch stall to nearly zero. DCPP also incurs 18.61 s of free idle time, whereas VPP reduces it to only 0.39 s. Overall, VPP reduces bubble ratio from 6.35\% to 0.14\%, achieving a 98.04\% reduction. This improvement comes from V-shaped scheduling, which mitigates inter-stage imbalance by filling otherwise idle windows with useful computation.
    
    \textbf{Exposed Communication.}
    Communication is the primary trade-off introduced by VPP. Compared with DCPP, VPP exposes 7.95\,s more communication on the critical path, mainly due to additional AllGather auxiliary overhead. Although VPP reduces the number of auxiliary AllGather calls by 4.58$\times$, the average latency per call increases by 11$\times$, leading to a 39.7\% (5.27\,s) increase in the overall AllGather overhead. However, VPP substantially improves communication hiding: the amount of communication overlapped with computation increases from 0.01\,s to 32.86\,s, raising the overlap ratio from 0.05\% to 51.75\%.

    \textbf{Impact of Sparse Attention.}
    We further profile GLM-5.2 on the 128K workload to understand how VPP interacts with sparse attention (DSA). Unlike Qwen and DeepSeek, where attention computation increases more significantly with the accumulated prefix context, DSA reduces the growth of attention cost at long contexts. As a result, the near-linear chunk latency scaling assumed by VPP no longer strictly holds: early chunks still benefit from V-shaped interleaving, while the scheduling efficiency gradually degrades as the sequence length increases, leading to higher per-chunk overhead and the re-emergence of bubbles in later chunks. Despite this limitation, VPP still achieves positive gains over both CPP and DCPP at contexts below 512K (Section~\ref{sec:Evaluation.End2End}), demonstrating that virtual-stage scheduling remains effective even with sparse attention.

\subsection{Ablation Study}
\label{sec:Evaluation.Ablation}
 
    \begin{figure}[t]
      \centering
      \includegraphics[width=\linewidth]{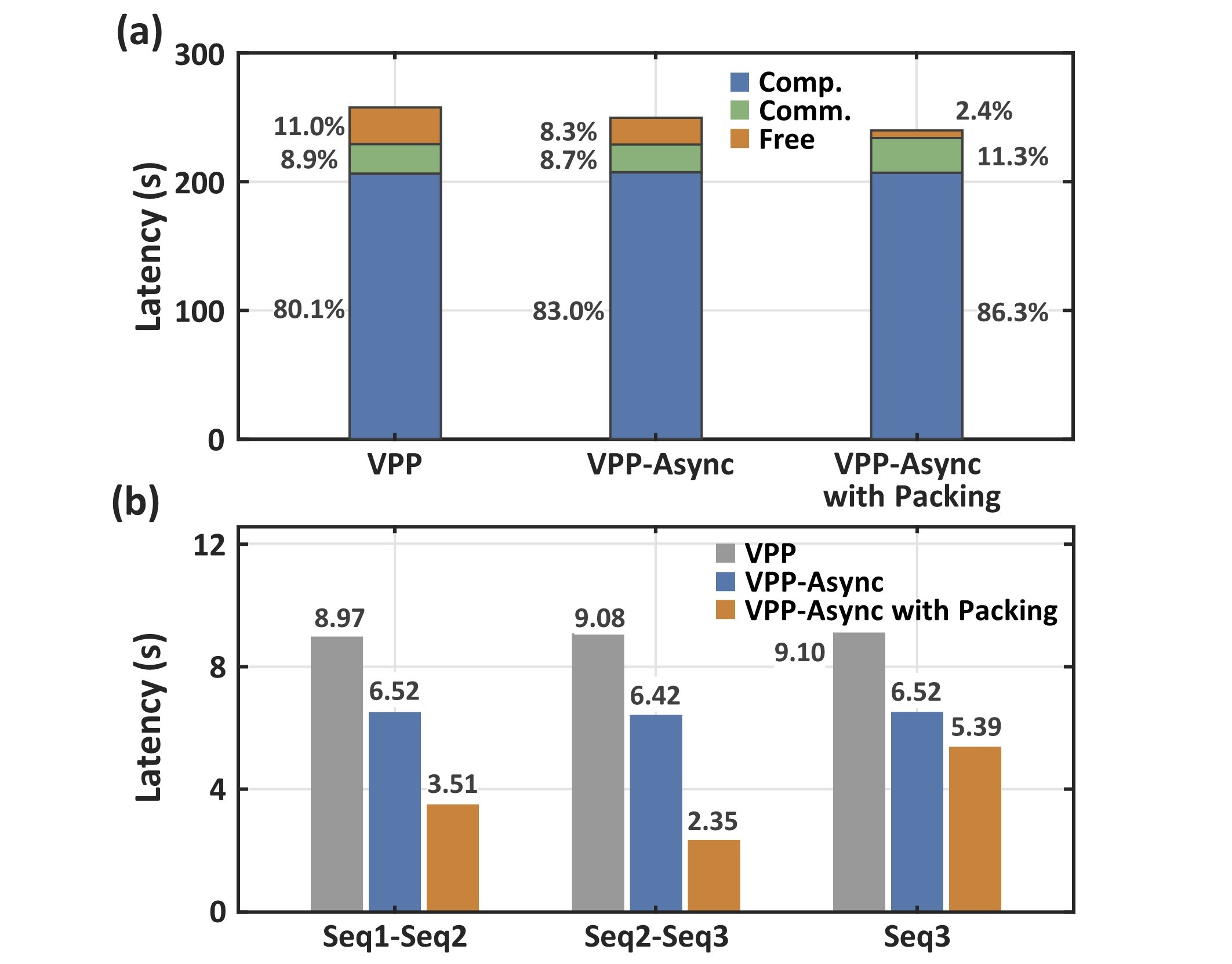}
      \caption{End-to-end performance of three VPP-based schemes. (a) Latency decomposed into computation (Comp.), exposed communication (Comm.), and Free. (b) Cross-request bubble latency.}
      \label{fig:3VPPE2EPerf}
    \end{figure}

    \begin{figure*}[t]
      \centering
      \includegraphics[width=\linewidth]{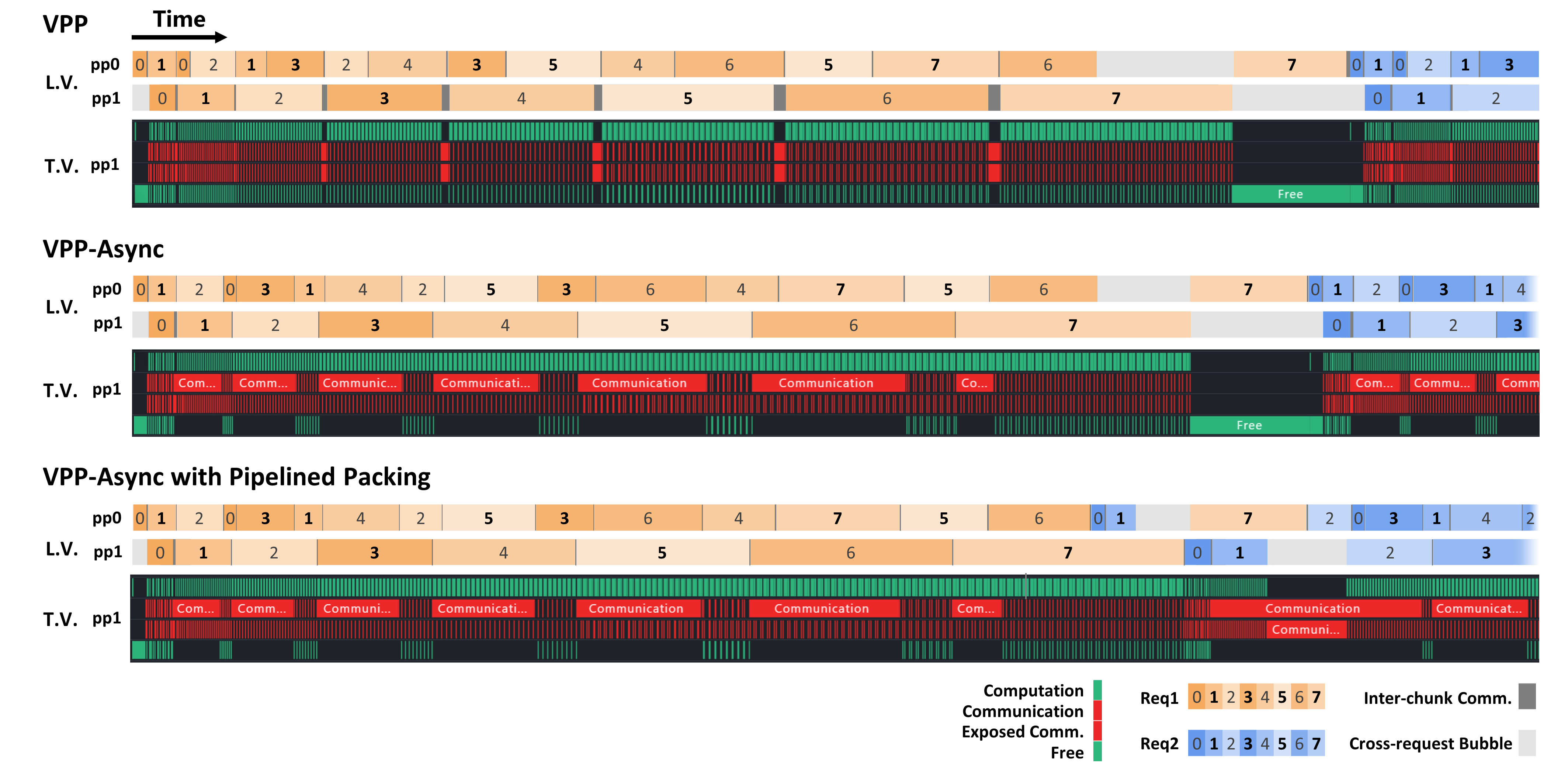}
      \caption{Execution timelines of two consecutive requests for the three VPP schemes. Logic View (L.V.): chunk-scheduling diagram derived from profiler traces. Timeline View (T.V.): raw profiler traces of computation and communication stream activity on $pp_1$.}
      \label{fig:VPPTimeline}
    \end{figure*}

    \begin{figure*}[t]
      \centering
      \includegraphics[width=\linewidth]{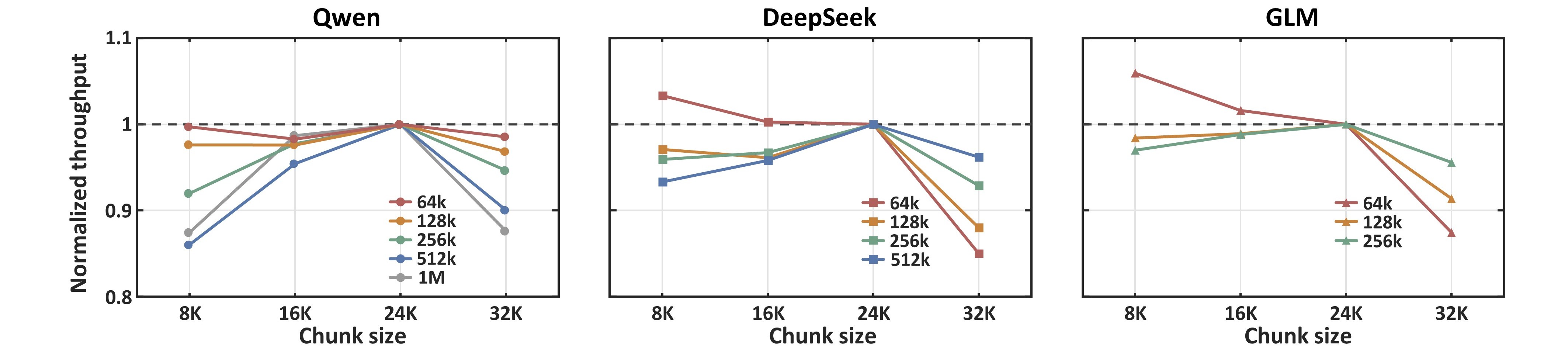}
      \caption{Normalized throughput for different chunk sizes across sequence lengths on three models, normalized to 24K.}
      \label{fig:ChunkSize}
    \end{figure*}

    We evaluate the two key design aspects of VPP: the contribution of each optimization stage and the impact of chunk size selection on long-sequence performance.
    
    \textbf{Ablation of VPP Variants.}
    We compare the three VPP variants in Figure~\ref{fig:VPPFramework}—vanilla VPP, VPP-Async, and VPP-Async with pipelined packing—to isolate the contribution of each component. This experiment uses a 256K-token workload on DeepSeek-V3.1 with a 32K chunk size and three concurrent requests. 
    
    Figure~\ref{fig:3VPPE2EPerf}(a) presents the end-to-end latency and its breakdown on the critical path ($pp_0$) for the three variants. Compared to vanilla VPP, VPP-Async reduces end-to-end latency by 3.13\%, and pipelined packing further improves the reduction to 6.89\%. Compute time differs by at most 0.82~s (0.42\%) across variants, confirming that the gains stem almost entirely from improved pipeline utilization rather than faster kernel execution. The major improvements come from reducing pipeline bubbles: the bubble ratio decreases from 11.0\% in vanilla VPP to 8.3\% with VPP-Async and further to 2.4\% with pipelined packing, mainly by eliminating cross-request idle periods, as shown in Figure~\ref{fig:3VPPE2EPerf}(b).

    Figure~\ref{fig:VPPTimeline} visualizes their chunk layouts and execution traces, denoted as Logic View (L.V.) and Timeline View (T.V.), respectively. The three variants isolate two distinct inefficiencies.  First, compared to vanilla VPP, VPP-Async overlaps communication with computation, reducing non-overlapped communication by 48\% and raising the communication overlap ratio from 0.02\% to 54.03\%. Second, pipelined packing fills the idle pipeline windows at request boundaries with the next request's leading chunks, reducing cross-request bubble latency by 33.2\% in this workload.

    \textbf{Chunk Size Sweep.} 
    Figure~\ref{fig:ChunkSize} sweeps the chunk size (8K--32K) for sequence lengths from 64K to 1M on three models, normalized to chunk size of 24K.
    
    The sweep reveals a counter-intuitive pattern: 24K performs best in most configurations because it does not divide sequences evenly. The final, smaller chunk executes during pipeline drain, overlapping its computation and communication with preceding chunks and reducing the tail bubble that an even split would leave. The only exception is 64K sequences on DeepSeek and GLM, where throughput drops monotonically with chunk size; with only a few chunks, the pipeline never reaches steady state, and smaller chunks directly shorten the dominant tail bubble.
    
% =========================================================
\section{Conclusion And Discussion}
\label{sec:Conclusion}
% =========================================================

    In this study, we presented VPP, a virtual pipeline parallelism design for efficient chunked prefill in long-context LLM inference. Rather than resizing chunks to fit the pipeline, VPP keeps chunk sizes fixed and optimizes the pipeline layout through virtual stages. V-shaped stage traversal exploits the approximately linear latency growth of causal attention to closely match computation across pipeline ranks, while asynchronous communication reordering and pipelined packing further reduce exposed communication stalls and cross-request drain bubbles. 
    
    Implemented on vLLM-Ascend, VPP delivers consistent throughput and TTFT gains across three MoE models and diverse workloads, improving throughput by up to 13.1\% over DCPP on long sequences and 6.7\% on mixed workloads while preserving performance on short sequences. On a 512K-token DeepSeek-V3.1 prefill workload, VPP reduces the pipeline bubble ratio from 6.4\% to 0.1\%, achieving a 98.0\% reduction compared with DCPP. 
    
    Beyond these improvements, our study highlights an important consideration for future pipeline optimization: the effectiveness of layout-based scheduling depends on the regularity of chunk latency growth. While dense attention exhibits predictable latency scaling that VPP can exploit, sparse attention reduces this regularity and limits the achievable bubble reduction. Extending virtual pipeline layouts to sparse-attention models, deeper pipelines, and more heterogeneous serving scenarios remains an important direction for future work.

% % Acknowledgements should only appear in the accepted version.
% \section*{Acknowledgements}


\bibliography{example_paper}
\bibliographystyle{mlsys2026}

%%%%%%%%%%%%%%%%%%%%%%%%%%%%%%%%%%%%%%%%%%%%%%%%%%%%%%%%%%%%%%%%%%%%%%%%%%%%%%%
%%%%%%%%%%%%%%%%%%%%%%%%%%%%%%%%%%%%%%%%%%%%%%%%%%%%%%%%%%%%%%%%%%%%%%%%%%%%%%%
% SUPPLEMENTAL CONTENT AS APPENDIX AFTER REFERENCES
%%%%%%%%%%%%%%%%%%%%%%%%%%%%%%%%%%%%%%%%%%%%%%%%%%%%%%%%%%%%%%%%%%%%%%%%%%%%%%%
%%%%%%%%%%%%%%%%%%%%%%%%%%%%%%%%%%%%%%%%%%%%%%%%%%%%%%%%%%%%%%%%%%%%%%%%%%%%%%%
\appendix
\setcounter{figure}{0}
\setcounter{table}{0}
\renewcommand{\thefigure}{A\arabic{figure}}
\renewcommand{\thetable}{A\arabic{table}}

\end{document}